\documentclass{sig-alternate-05-2015}
\usepackage[utf8]{inputenc}
\usepackage{hyperref}
\usepackage{algorithm,algorithmicx}
\usepackage{amsfonts,amsmath, amssymb, bm}
\usepackage[numbers]{natbib}% citations
\usepackage{times}
\usepackage{helvet}
\usepackage{courier}
\usepackage{url}
\usepackage{hyperref}
\usepackage{amsmath}
\usepackage{epsfig}
\usepackage{subfigure}
\makeatletter
\def\@maketitle{\newpage
 \null
 \setbox\@acmtitlebox\vbox{%
\baselineskip 20pt
\vskip 2em                   % Vertical space above title.
   \begin{center}
    {\ttlfnt \@title\par}       % Title set in 18pt Helvetica (Arial) bold size.
    \vskip 1.5em                % Vertical space after title.
{\subttlfnt \the\subtitletext\par}\vskip 1.25em%\fi
    {\baselineskip 16pt\aufnt   % each author set in \12 pt Arial, in a
     \lineskip .5em             % tabular environment
     \begin{tabular}[t]{c}\@author
     \end{tabular}\par}
    \vskip 0.5em               % Vertical space after author.
   \end{center}}
 \dimen0=\ht\@acmtitlebox
 \unvbox\@acmtitlebox
 \ifdim\dimen0<0.0pt\relax\vskip-\dimen0\fi}
\makeatother

\makeatletter
\def\@copyrightspace{\relax}
\makeatother

\begin{document}

% Copyright
%\setcopyright{acmcopyright}
%\setcopyright{acmlicensed}
%\setcopyright{rightsretained}
%\setcopyright{usgov}
%\setcopyright{usgovmixed}
%\setcopyright{cagov}
%\setcopyright{cagovmixed}

% DOI
%\doi{10.475/123_4}

% ISBN
%\isbn{123-4567-24-567/08/06}

%Conference
%\conferenceinfo{PLDI '13}{June 16--19, 2013, Seattle, WA, USA}

%\acmPrice{\$15.00}

%
% --- Author Metadata here ---
%\conferenceinfo{WOODSTOCK}{'97 El Paso, Texas USA}
%\CopyrightYear{2007} % Allows default copyright year (20XX) to be over-ridden - IF NEED BE.
%\crdata{0-12345-67-8/90/01}  % Allows default copyright data (0-89791-88-6/97/05) to be over-ridden - IF NEED BE.
% --- End of Author Metadata ---

\title{High Performance Scalable FPGA Accelerator for Deep Neural Networks}
\author{
  Sudarshan Srinivasan, Pradeep Janedula, Saurabh Dhoble, Sasikanth Avancha, Dipankar Das, \\
 Naveen Mellempudi, Bharat Daga, Martin Langhammer, Gregg Baeckler, Bharat Kaul\\ 
    {Intel Corporation}\\
     {\{sudarshan.srinivasan\}@intel.com}
}

\maketitle
%\begin{abstract}

%\end{abstract}
\section{abstract}\label{sec:abstract}
Low-precision is the first order knob for achieving higher Artificial Intelligence Operations (AI-TOPS). However the algorithmic space for sub-8-bit precision compute is diverse, with disruptive changes happening frequently, making FPGAs a natural choice for Deep Neural Network inference, In this work we present an FPGA-based accelerator for CNN inference acceleration. We use {\it INT-8-2} compute (with {\it 8 bit} activation and {2 bit} weights) which is recently showing promise in the literature, and which no known ASIC, CPU or GPU natively supports today. Using a novel Adaptive Logic Module (ALM) based design, as a departure from traditional DSP based designs, we are able to achieve high performance measurement of 5 AI-TOPS for {\it Arria10} and project a performance of 76 AI-TOPS at 0.7 TOPS/W for {\it Stratix10}. This exceeds known CPU, GPU performance and comes close to best known ASIC (TPU) numbers, while retaining the versatility of the FPGA platform for other applications. 
\def\argmin{\mathop{\hbox{argmin}}\limits}
\def\argmax{\mathop{\hbox{argmax}}\limits}

\def\mbf#1{{\mbox{\boldmath{$#1$}}}}
\def\x{{\mathbf x}}
\def\y{{\mathbf y}}
\def\z{{\mathbf z}}
\def\X{{\mathbf X}}
\def\Y{{\mathbf Y}}
\def\a{{\mathbf a}}
\def\b{{\mathbf b}}
%%---- \def\exp#1{{\left\langle#1\right\rangle}}
\def\E#1{{{\mathbf E}\left[#1\right]}}
%\def\mod#1{{\left|#1\right|}}
%%---- \def\norm#1{{\|#1\|}}
%%---- \def\Norm#1{{\left\|#1\right\|}}
\def\floor#1{{\left\lfloor\,#1\,\right\rfloor}}
\def\ceil#1{{\left\lceil\,#1\,\right\rceil}}
\def\r#1{{(\ref{#1})}}
\def\rot#1{{\rotatebox{90}{#1\ }}}

\newcommand{\Probab}[1]{\mbox{}{\bf{Pr}}\left[#1\right]}
\newcommand{\Expect}[1]{\mbox{}{\bf{E}}\left[#1\right]}
\newcommand{\Varnce}[1]{\mbox{}{\bf{Var}}\left[#1\right]}
\newcommand{\Trace }[1]{\mbox{}{\bf{Tr}}\left(#1\right)}
\newcommand{\Sqrt  }[1]{\mbox{}\left(#1\right)^{1/2}}
\newcommand{\Qdrt  }[1]{\mbox{}\left(#1\right)^{1/4}}
\newcommand{\sNorm }[1]{\mbox{}\|#1\|  }
\newcommand{\ONorm }[1]{\mbox{}\left\|#1\right\|_{\ell_1}  }
\newcommand{\OsNorm }[1]{\mbox{}\|#1\|_{\ell_1}  }
\newcommand{\ONormS}[1]{\mbox{}\left\|#1\right\|_{\ell_1}^2}
\newcommand{\ZNorm }[1]{\mbox{}\left\|#1\right\|_0  }
\newcommand{\ZsNorm }[1]{\mbox{}\|#1\|_0  }
\newcommand{\ZNormS}[1]{\mbox{}\left\|#1\right\|_0^2}
\newcommand{\FsNorm }[1]{\mbox{}\|#1\|_F  }
\newcommand{\FNorm }[1]{\mbox{}\left\|#1\right\|_F  }
\newcommand{\FNormS}[1]{\mbox{}\left\|#1\right\|_F^2}
\newcommand{\FNormQ}[1]{\mbox{}\left\|#1\right\|_F^4}
\newcommand{\TNorm}[1]{\mbox{}\left\|#1\right\|_2}
\newcommand{\TsNorm}[1]{\mbox{}\|#1\|_2}
\newcommand{\TNormS}[1]{\mbox{}\left\|#1\right\|_2^2}
\newcommand{\TNormQ}[1]{\mbox{}\left\|#1\right\|_2^4}
\newcommand{\XNorm }[1]{\mbox{}\left\|#1\right\|_{\xi}  }
\newcommand{\XNormS}[1]{\mbox{}\left\|#1\right\|_{\xi}^2}
\newcommand{\XNormQ}[1]{\mbox{}\left\|#1\right\|_{\xi}^4}
\newcommand{\PNorm }[1]{\mbox{}\left\|#1\right\|_p  }
\newcommand{\PNormS}[1]{\mbox{}\left\|#1\right\|_p^2}
\newcommand{\VTNorm }[1]{\mbox{}\left|#1\right|  }
\newcommand{\VTNormS}[1]{\mbox{}\left|#1\right|^2}
\newcommand{\VTNormQ}[1]{\mbox{}\left|#1\right|^4}
\newcommand{\VTTNorm }[1]{\mbox{}\left\|#1\right\|_2  }
\newcommand{\VTTNormS}[1]{\mbox{}\left\|#1\right\|_2^2}
\newcommand{\VTTNormQ}[1]{\mbox{}\left\|#1\right\|_2^4}
\newcommand{\VINorm }[1]{\mbox{}\left\|#1\right\|_{\infty}  }
\newcommand{\VIsNorm }[1]{\mbox{}\|#1\|_{\infty}  }
\newcommand{\VINormS}[1]{\mbox{}\left\|#1\right\|_{\infty}^2}
\newcommand{\VINormQ}[1]{\mbox{}\left\|#1\right\|_{\infty}^4}
\newcommand{\MaxNorm }[1]{\mbox{}\|#1\|_{\max}  }

\newcommand{\setlinespacing}[1]%
           {\setlength{\baselineskip}{#1 \defbaselineskip}}
\newcommand{\doublespacing}{\setlength{\baselineskip}%
                           {2.0 \defbaselineskip}}
\newcommand{\singlespacing}{\setlength{\baselineskip}{\defbaselineskip}}

\newcommand{\inner }[1]{\left<#1\right>}
\newcommand{\rank}[1]{{\bf rank}{\left(#1\right)}}
\newcommand{\abs }[1]{\left|#1\right|}
\newcommand{\sabs }[1]{|#1|}
\newcommand{\absS}[1]{\left|#1\right|^{2}}
\newcommand{\spann}[1]{\mbox{\bf span}\left(#1\right)}
\newcommand{\mcut}[1]{\mbox{\bf MAX-CUT}\left[#1\right]}
\newcommand{\qe}{\hfill  \rule{2mm}{3mm}}

\newtheorem{definition}{Definition}
\newtheorem{proposition}{Proposition}
\newtheorem{lemma}{Lemma}
\newtheorem{theorem}{Theorem}
\newtheorem{corollary}{Corollary}
\newtheorem{problem}{Problem}
\newenvironment{Proof}{\noindent {\em Proof:}}{\hspace*{\fill}\mbox{$\diamond$}}

\newcommand{\mat}[1]{{\ensuremath{\bm{\mathrm{#1}}}}}
\def\gammab{{\bm{\gamma}}}
\def\kappab{{\bm{\kappa}}}
\def\sig{{\bm{\Sigma}}}
\def\sigplus{{\bm{\Sigma}^{+}}}
\def\siginv{{\bm{\Sigma}^{-1}}}
\def\bet{{\bm{\beta}}}
\def\one{{\bm{1}}}
\def\exp{\hbox{\rm exp}}
\def\rank{\hbox{\rm rank}}
\def\col{\hbox{\rm col}}
\def\ker{\hbox{\rm ker}}
\def\ahat{{\hat\a}}
\def\a{{\bm \alpha}}
\def\w{{\mathbf w}}
\def\hw{{\mathbf{\hat{w}}}}
\def\ta{\tilde{\bm \alpha}}
\def\tw{\tilde{\mathbf w}}
\def\b{{\mathbf b}}
\def\e{{\mathbf e}}
\def\expe{{\mathrm e}}
\def\q{{\mathbf q}}
\def\rb{{\mathbf r}}
\def\s{{\mathbf s}}
\def\x{{\mathbf x}}
\def\y{{\mathbf y}}
\def\z{{\mathbf z}}
\def\u{{\mathbf u}}
\def\v{{\mathbf v}}
\def\d{{\mathbf \delta}}
\def\xhat{{\hat\x}}
\def\yhat{{\hat\y}}
\def\A{\matA}
\def\B{\matB}
\def\C{\matC}
\def\Ahat{\hat\matA}
\def\Atilde{\tilde\matA}
\def\Btilde{\tilde\matB}
\def\Stilde{\tilde\matS}
\def\Utilde{\tilde\matU}
\def\Vtilde{\tilde\matV}
\def\E{{\cl E}}
\def\G{{\cl G}}
\def\hset{{\cl H}}
\def\Q{{\bm{Q}}}
\def\U{{\bm{U}}}
\def\V{{\bm{V}}}
\def\win{\hat{\w}}
\def\wopt{\w^*}
\def\matAhat{\hat\mat{A}}
\def\matA{\mat{A}}
\def\matB{\mat{B}}
\def\matC{\mat{C}}
\def\matD{\mat{D}}
\def\matE{\mat{E}}
\def\matG{\mat{G}}
\def\matH{\mat{H}}
\def\matI{\mat{I}}
\def\matK{\mat{K}}
\def\matM{\mat{M}}
\def\matP{\mat{P}}
\def\matQ{\mat{Q}}
\def\matR{\mat{R}}
\def\matRhad{\matR_{\textsc{\tiny srht}}}
\def\matS{\mat{S}}
\def\matT{\mat{T}}
\def\matU{\mat{U}}
\def\matV{\mat{V}}
\def\matW{\mat{W}}
\def\matX{\mat{X}}
\def\matY{\mat{Y}}
\def\matZ{\mat{Z}}
\def\matSig{\mat{\Sigma}}
\def\matDelta{\mat{\Delta}}
\def\matPhi{\mat{\Phi}}
\def\matTh{\mat{\Theta}}
\def\matGam{\mat{\Gamma}}
\def\matPsi{\mat{\Psi}}
\def\vv{\mathbf{v}}
\def\Exp{\mathbb{E}}
% ==============

%\def\R{{\mathbf R\hspace*{-1.65ex}\rule{0.08ex}{1.35ex}\hspace*{1.65ex}}}
\def\Nat{{\mathbf N\hspace*{-1.89ex}\rule{0.07ex}{1.5ex}\hspace*{1.89ex}\hspace*{-1.7ex}\rule{0.1ex}{1.5ex}\hspace*{1.7ex}\hspace*{-.6ex}\rule{0.1ex}{1.5ex}\hspace*{.6ex}}}
%\def\Q{{\mathbf Q\hspace*{-1.35ex}\raise.15ex\hbox{\rule{0.095ex}{1.35ex}}\hspace*{1.4ex}}}
%\def\C{{\mathbf C\hspace*{-1.25ex}\raise.15ex\hbox{\rule{0.095ex}{1.35ex}}\hspace*{1.25ex}}}
%MWM% \def\qed{\hfill\rule{2mm}{2mm}}
\def\sm{\sum\limits}
\def\choose#1#2{\left({{#1}\atop{#2}}\right)}
\def\cl#1{{\cal #1}}
\def\prob#1{{P\left[{#1}\right]}}
\def\mathbox#1{{\fbox{\math{#1}}}}
\def\mandbox#1{{\fbox{\math{\displaystyle #1}}}}
\def\argmin{\mathop{\hbox{argmin}}\limits}
\def\argmax{\mathop{\hbox{argmax}}\limits}

\def\mbf#1{{\mbox{\boldmath{$#1$}}}}
\def\x{{\mathbf x}}
\def\y{{\mathbf y}}
\def\z{{\mathbf z}}
\def\X{{\mathbf X}}
\def\Y{{\mathbf Y}}
\def\a{{\mathbf a}}
\def\b{{\mathbf b}}
%%---- \def\exp#1{{\left\langle#1\right\rangle}}
\def\E#1{{{\mathbf E}\left[#1\right]}}
%\def\mod#1{{\left|#1\right|}}
%%---- \def\norm#1{{\|#1\|}}
%%---- \def\Norm#1{{\left\|#1\right\|}}
\def\floor#1{{\left\lfloor\,#1\,\right\rfloor}}
\def\ceil#1{{\left\lceil\,#1\,\right\rceil}}
\def\r#1{{(\ref{#1})}}
\def\rot#1{{\rotatebox{90}{#1\ }}}

\newcommand{\stitle}[1]{\begin{center}{\bf\color[named]{Red}#1}\\[-.8em]\rule{\li
newidth}{1pt}\end{center}}
\newcommand{\myv}[1]{{ \bf #1} }
\newcommand{\mym}[1]{{ \bf #1} }
\newcommand{\red}[1]{{\color[named]{Red} #1}}
\newcommand{\blue}[1]{{\color[named]{Blue} #1}}
\newcommand{\green}[1]{{\color[named]{Green} #1}}
\newcommand{\light}[1]{{\color[rgb]{0.7,0.7,0.7} #1}}
\newcommand{\dark}[1]{{\color[rgb]{0,0,0} #1}}

\def\qn#1{{{#1}?\hfill\bf\math{\bigcirc} 1\qquad\math{\bigcirc} 2\qquad\math{\bigcirc} 3\qquad\math{\bigcirc} 4\qquad\math{\bigcirc} 5}}

\def\dotfil{\leaders\hbox to 1.5mm{.}\hfill}
\newcounter{rmnum}
\def\RN#1{\setcounter{rmnum}{#1}\uppercase\expandafter{\romannumeral\value{rmnum}}}
\def\rn#1{\setcounter{rmnum}{#1}\expandafter{\romannumeral\value{rmnum}}}
\def\lequal{{\buildrel{\scriptscriptstyle <}\over{\scriptscriptstyle =}}}
\def\gequal{\ {\buildrel{\scriptstyle >}\over{\scriptstyle =}}\ }
\def\econd#1#2{\ {\buildrel{\scriptstyle #1}\over{\scriptstyle #2}}\ }

\def\smileyface{{
\begin{picture}(13,13)
\allinethickness{1.5pt}
\put(6.5,3.5){\circle{13}}
\put(3.9,5.5){\circle*{1.25}}
\put(8.7,5.5){\circle*{1.25}}
\put(1.7,0){\spline(2,2.5)(4.7,0)(7,2.5)}
\end{picture}
}}
\def\sadface{{
\begin{picture}(13,13)
\allinethickness{1.5pt}
\put(6.5,3.5){\circle{13}}
\put(3.9,5.5){\circle*{1.25}}
\put(8.7,5.5){\circle*{1.25}}
\put(1.7,0.5){\spline(2.5,0)(5,2.5)(7.5,0)}
\end{picture}
}}

\vspace{-0.5em}
\section{Introduction}\label{sec:Introduction}
Deep Neural Networks (DNNs) have achieved state-of-the-art accuracy for a variety of inference tasks in various domains such as object detection, image classification, and speech recognition. However, this is achieved at a very high computation (~8 GFlop per image classification using ResNet50), memory footprint (~100 MB), and bandwidth \cite{ResNet}. 
One way to circumvent this problem is to perform low-precision compute \cite{TernaryFGQ, CourbariauxB16, HubaraCSEB16, HeZRS15, GuptaAGN15}, so much so that {\it 8-bit} compute has today become the mainstay for inference \cite{google}, yielding 4x or higher benefit over {\it FP32}. 
At the same time, research is pushing the boundaries even upto ternary (-1, 0, +1), and binary operations, thereby significantly boosting the AI-TOPs (Artifitial Intellegence Tera Operations per second) - loosely defined as the amount of lowest precision compute at which acceptably accurate inference can be performed by a machine.

However, as we stride into the 8-bit and sub-8 bit compute domain to leverage higher performance, we discover significantly higher diversity amongst options for low precision compute. These not only vary in specific details of the algorithm used (for example GEMMLOWP vs Dynamic Fixed Point), but also in the type of data used ({\it FP or INT}) and precision of activations, and weights.  Indeed one finds across the literature almost all combinations of precision-pairs for activations and weights (for example {\it 8 bit} activation {\it 2 bit} weights \cite{TernaryFGQ}, {\it 5 bit} activation, and {\it 4 bit} weights. Moreover, the precision of operations is dependent on the application and sometimes varies even between different operations (or layers) in the application leading to a hybrid precision scenario.     
Such being the diversiry of low-precision inference for DNNs, it requires a thorough investigation of FPGA's for DNN inference, where the precision can be arbitrarily altered for activations and weights across applications and parts of an application. Not surprising then is the wide array of FPGA offerings for 8-bit and sub-8-bit inference \cite{Nurvitadhi:2017, Duncan, BNN}. It may also be noted that while ASIC's will always have higher compute density than FPGA's, the fast evolving algorithmic landscape makes things challenging for ASIC's, while on the other hand FPGA's can be used for multiple different (non-DNN) applications, thereby providing a much higher value proposition. 

An interesting {\it sub-8-bit} design point emerges when we have integer {\it 8-bit} activations and ternary weights. Here, the activations can fit into on-chip memory or many modern FPGA's like the Arria-10 \cite{Arria10}. For inference, we can swap between input and output buffers for consecutive layers, thereby needing a memory footprint of 2-3 buffers for storing activation of one data point. At the same time, weights can stream from memory at almost the lowest achievable bandwidth (lowest achieved for binary weights), or can even reside in memory for certain cases. This design choice is further buttressed by competitive accuracy results for state-of-the-art Convolutional Neural Networks (like ResNet50) on ImageNet-1k dataset using {\it INT8-2} and fine grained quantization (FGQ) \cite{TernaryFGQ}. FGQ using {\it INT-8-2} achieves 4\% lower Top-1 accuracy.
%while this technique is extended to Residual Ternary networks to obtain SOTA accuracy matching {\it FP32} results \cite{residual ternary}, while {\it INT-8-2} is also used in \cite{AAL paper} to achieve results which are x\% off-SOTA.     

From an FPGA design persective an {\it INT-8-2} design provides an additional interesting opportunity. Since ternary weights are used, all MAC (Multiply-and-Accumulate) operations are essentially replaced by additions. This allows us to explore the use of Adaptive Logic Modules (ALM's) for the MAC operations, in contrast to the traditional use of DSP's for Neural Network Compute. This is critical as FPGA's typically have equivalent or less density of DSP Flops \cite{Arria10}, as compared to CPU's, or GPU's \cite{Nvidia}. Moreover a majority of an FPGA floorplan consists of ALM's and not DSP's, so leveraging ALM's for DNN compute becomes critical to competitive AI-TOPs with respect to CPU's and GPU's.     

In this work we present an FPGA design for {\it INT-8-2} computation using the Fine Grained Quantization (FGQ) method. We design IP components which are tailored to the FGQ method to minimize algorithmic overheads, and a systolic array based Processing Element optimized for {\it INT-8-2}. Data locality is re-used to fully maximize the efficiency of the design. The main contributions of this work are:  
\noindent $\bullet$ To the best of our knowledge, this is the first work which uses ALM's for end-to-end NN compute. We conduct our experiments with ResNet50.

\noindent $\bullet$ A FPGA design co-optimized for the low-precision algorithm.

\noindent $\bullet$ Power performance projection of {\it 0.7 AI-TOPS/W} for inference on the Stratix-10, exceeding that of any currently available CPU, or GPU and closely matching ASIC's like TPU (1.2 TOPS/W).

\vspace{-1.0em}
\section{Related Work}\label{sec:Related Work}
Related work can be summarized into two main categories.  The first category focuses on recent research that improves accuracy of DNN's with low precision data types.  The second focuses on efficient hardware implementation of low precision DNN's.

\subsection{Accuracy of Low Precision DNN}
DNN's inferencing with low precision data types are a well researched topic \cite{TernaryFGQ}.  Many prior works have investigated low precision for weights while keeping activations at full precision\cite{Lin2015, RastegariORF16}.  Stochastic binarization scheme was shown to achieve state-of-the-art (SOTA) accuracies on smaller data sets (MNIST, CIFAR10, SVHN) in \cite{Lin2015}.  Near SOTA accuracy with AlexNet Topology on a larger ImageNet data set was shown in \cite{RastegariORF16}.  the above works have retained activations at full precision.  However, to realize full power/potential benefit for low precision weights, activations also need to be at lower precision.  Researchers have also demonstrated that the use of 8-4 bits shows reasonable high accuracy compared to full precision \cite{Philip}.
 
Binary Neural Network (BNN) investigates the use of 1-bit values for the weights, where weights are constrained to either +1 or -1 \cite{CourbariauxB16}.  BNN's have shown to achieve SOTA accuracy on smaller data sets (ex., CIFAR10).  XNOR-Net, which use binarized weights and activations, achieves top1-accuracy drop of 12\% for AlexNet \cite{RastegariORF16}.  Thus, achieving high accuracy with binary networks is still a challenge.  Ternary Neural Networks (TNN) investigate the use of 2-bit values for weights, where weights are constrained to +1,-1, and 0.  Top-1 error rate of 25.6\% has been reported for ResNet-50 models trained on ImageNet with 32-bit activations and ternary weights \cite{ZhuHMD16}.  Recently, Top-1 error of 29.24\% for ResNet 50 with 8-bit activations and ternary weights was achieved with the FGQ technique \cite{TernaryFGQ}.  As this is the highest accuracy number reported on ResNet with ImageNet for 8-2 thus far, we use the models (FGQ) presented in this work to further fine-tune the accuracy for ResNet network with ImageNet.

\subsection{Hardware Implementation of Low Precision DNN}
DNN can be efficiently mapped into a FPGA to achieve high performance as shown in various prior works \cite{Suda, Gysel, Zhang:2015}.  Several prior research works have implemented DNN's on a FPGA with precision width of 16/32-bits.  As a result, they use DSP for the core computation, failing to not make maximum use of the abundant logic resources available on modern FPGA's.  Very few works have evaluated DNN implementation for topologies like ResNet on modern FPGA such as the Stratix 10 to evaluate performance.  Stratix 10 performance for Ternery DNN on ResNet network was shown in \cite {Nurvitadhi:2017}.  However, the activations are still constrained to 32-bits, thus not fully utilizing the ALM's available.  ResNet implementation on the Arria10 FPGA with 16-bit precision was implemented in  \cite{ResNet}.  Maximum Tops that could be achieved by this design was only 0.28Tops.  This is the first work that implements 8-2 on a FPGA with dynamic fixed point support to achieve high performance for ResNet topology.

\vspace{-1.3em}
\section{Ternary ResNet Network} \label{sec:Ternary Residual Network}
 \vspace{-1.0em}
\subsection{ResNet Topology Overview}
\begin{figure}[h] \begin{center}
 \includegraphics[scale=0.35]{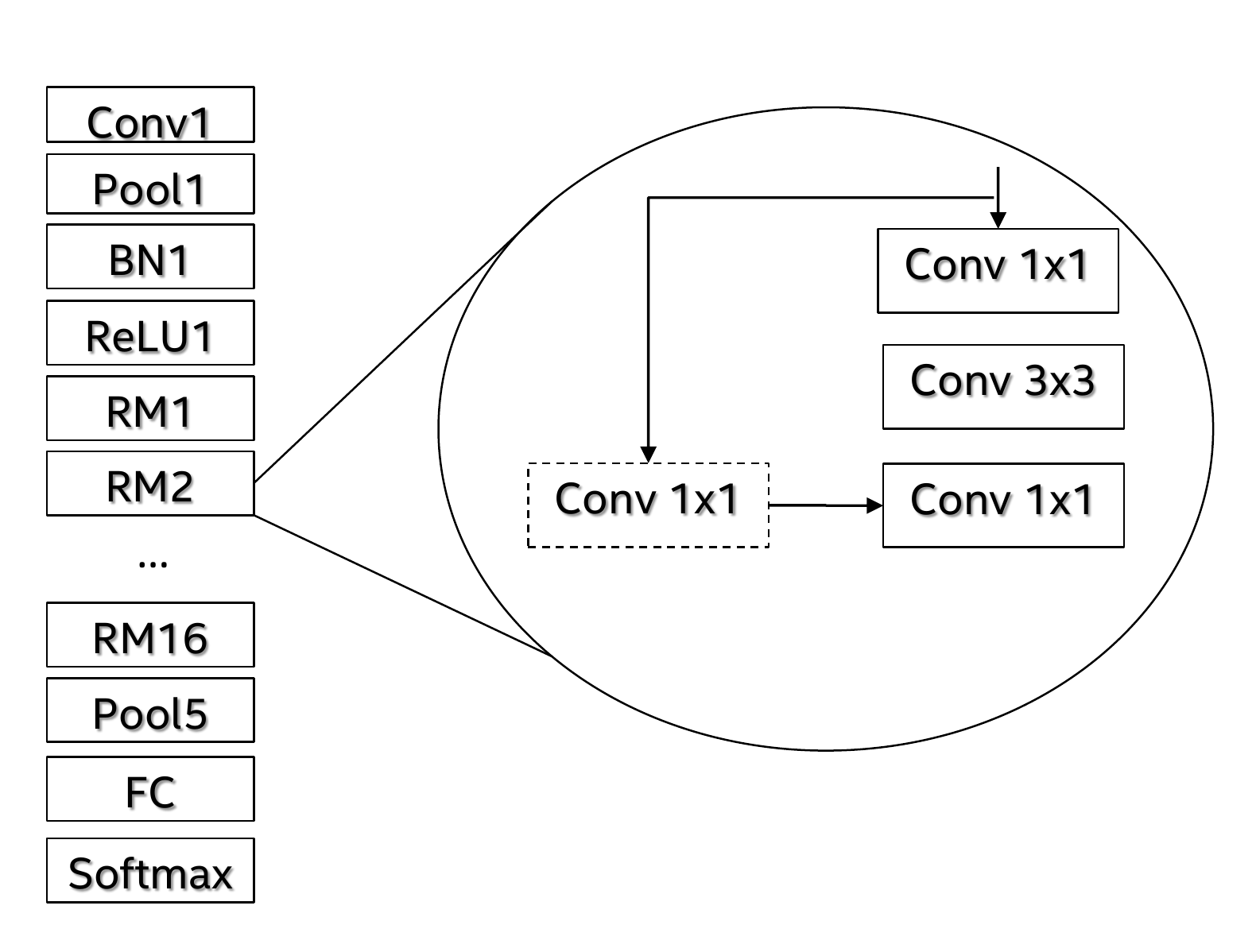}
 \end{center} \vspace{-2em}
 \caption{ResNet 50 Topology. RM denotes ResNet module}.
  \label{fig:ResNet} \vspace{-2em}
\end{figure}

\begin{figure}[h] \begin{center}
 \includegraphics[scale=0.35]{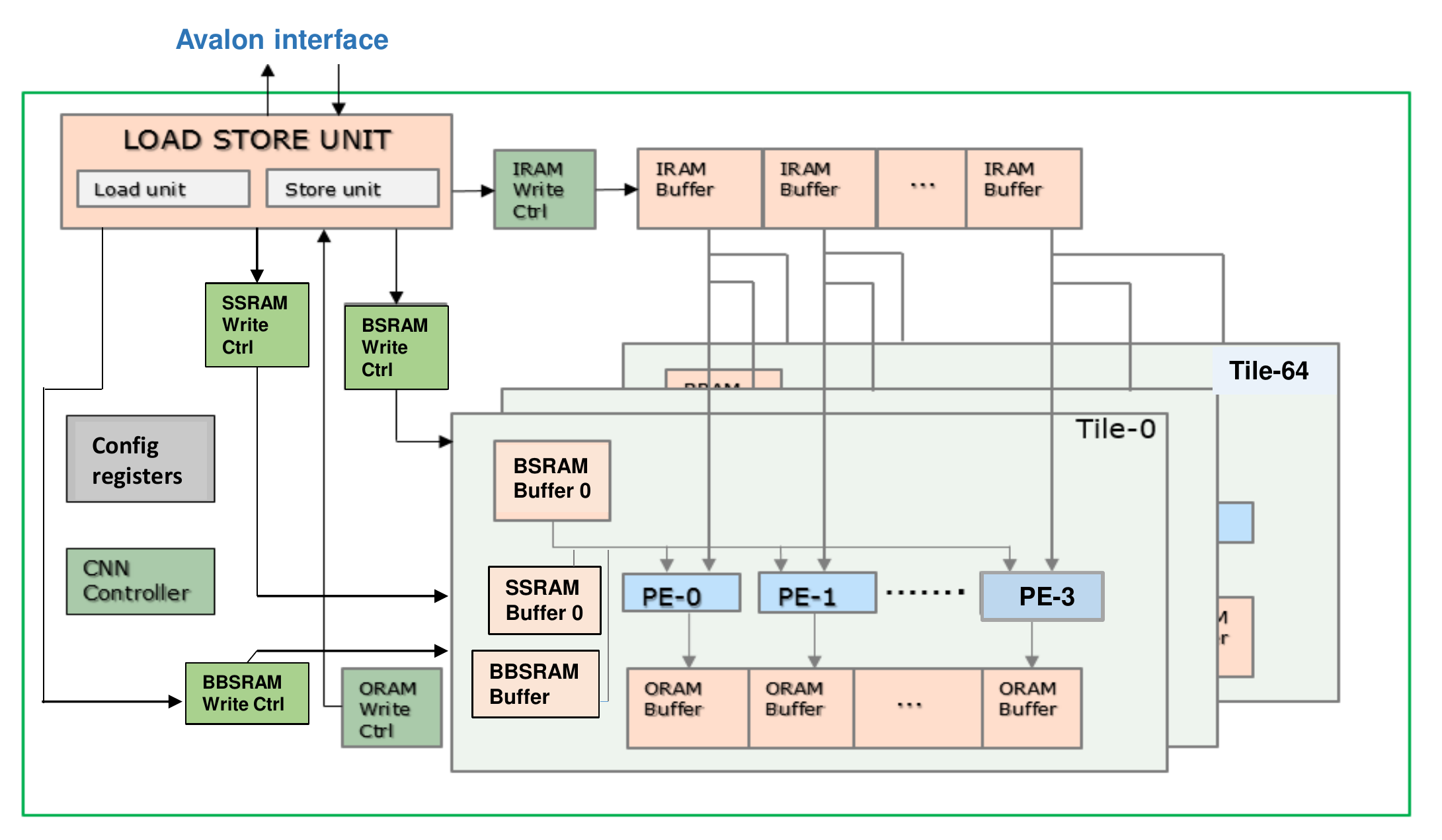}
 \end{center} \vspace{-2em}
 \caption{Overall system design of our FPGA accelerator contains various computing modules and memory elements.}
  \label{fig:System_design} \vspace{-2.0em}
\end{figure}

Deep residual networks (ResNet) have shown superior classification accuracy with fewer model parameters compared to previous DNN models \cite{Res2015}. ResNet network has become the benchmark in industry/academia for DNN's.  Efficient implementation of the network on hardware is important.  A typical ResNet topology consists of 16 different ResNet Modules (RM) with each consisting of convolution, max-pooling, batch norm, scale, and relu as shown Figure \ref{fig:ResNet}.  ResNet topologies have varying kernel dimensions across layers.  Thye also have skip connections, where the left and right branch of the ResNet are merged into an element-wise layer for summation.  The above characteristics of a ResNet have made the topology highly irregular compared to previous DNN models such as AlexNet, and VGG.  Therefore, it is highly challenging to implement ResNet topology on a FPGA to achieve high performance with fixed hardware and memory resources.

To achieve high accuracy on ImageNet data set using ResNet models, the first (Conv1, Pool1, BN1 on CPU) and last layers (Pool5, FC, Softmax), shown in Figure \ref{fig:ResNet}, are used at high precision with 8-bit activations(a) and 8-bit weights(w).  Our FPGA accelerator is designed to support only 8a-2w to make the design simpler and extract maximum performance from the FPGA.  Since the first and last layers work at high precision, they are run on the CPU.  As the other layers are implemented on the FPGA, they function only on convolutions, with the batch norm and scale layers fused with convolution filters.
 \vspace{-0.8em}
\subsection{ResNet-50 Model Optimization}
A FGQ based technique was applied on ResNet 50 network to achieve high accuracy for 8-bit activations and ternary weights \cite{TernaryFGQ}.  The fine grain quantization technique proposed divides the learned weights into sets of disjoint blocks (N), and then ternarizes them.  There is a scaling factor (alpha) associated with a block of size (N).  It is shown that for N = 64, 99\% of MAC operations can be replaced by ternary accumulations. 
This results in 15x potential improvement in performance \cite{TernaryFGQ}. For ResNet, the number of inputs and output channels scale by a multiple of 64. Thus we have taken N=64 in our work.
However, during inferencing, the batch norm and scale parameters in ResNet layers can be fused into the alpha scaling parameter, which was not considered in \cite{TernaryFGQ}. The fusion of these layers helps in easy implementation of the network on hardware.

Below, we outline how the fusion can be carried out with the FGQ technique in this work.
Let $\odot$ denote the ternary convolution operation. Let $\hat\matX$ denote the quantized activations. After fusing BN and scaling parameters, for a given block of FP32 weights $\matW^{(j)}$ and for a given ofm, we scale the FP32 weights by $\beta/\sigma$ with a bias of $\left(\gamma - \frac{\beta\cdot\mu}{\sigma}\right)$. Then, fused FP32 weights become
$$
\tilde\matW{}^{(j)} = \frac{\beta}{\sigma}\matW^{(j)}.
$$
We ternarize this $\tilde\matW{}^{(j)}$ as 
$\hat\alpha{}^{(j)}\hat\matW{}^{(j)}$, where $\hat\alpha{}^{(j)}$ is a quantized scaling factor and $\hat\matW{}^{(j)}_i\in\{-1,0,+1\}$, $\forall i$.
Then, the (partial) output of ternary convolution, for the given ofm, is
%Inference on ternary 8-2 setting
$$
\hat\y_j = (\hat\matX\odot \hat\matW{}^{(j)})\hat\alpha{}^{(j)}
+\left(\gamma - \frac{\beta\cdot\mu}{\sigma}\right),
$$
and the full output, for the given ofm, is
$$
\hat\y = \sum_j (\hat\matX\odot \hat\matW{}^{(j)})\hat\alpha{}^{(j)}
+ \left(\gamma - \frac{\beta\cdot\mu}{\sigma}\right).
$$

\section{Scalable FPGA DNN Architecture}\label{sec:RelatedWork}
 \vspace{-1em}
\subsection{Architecture Description}
\begin{figure}[h] \begin{center}
 \includegraphics[scale=0.45]{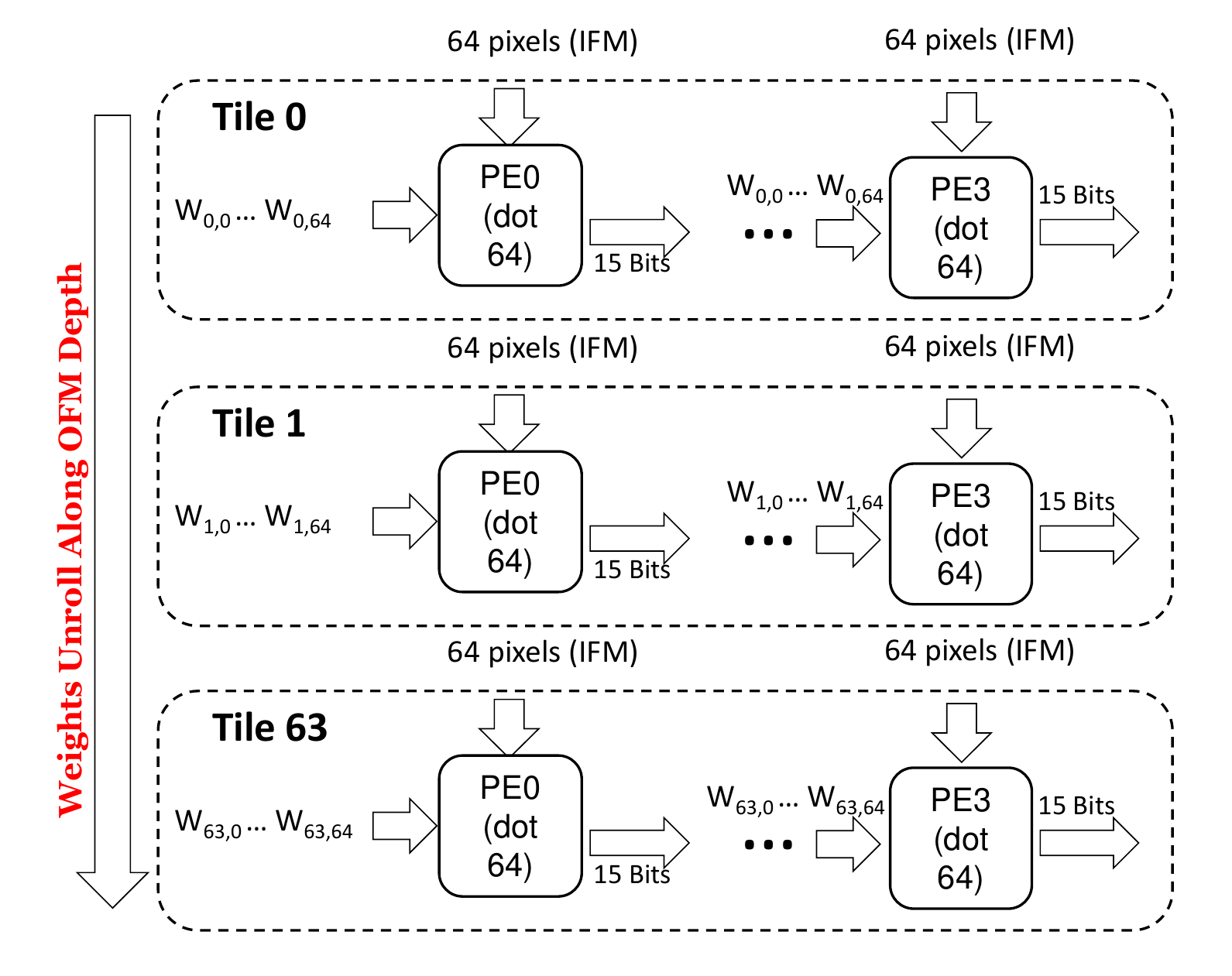}
 \end{center} \vspace{-3em}
 \caption{Tile layout of our architecture containing 4 PE per tile with weights and input fed into each of the PE}
  \label{fig:Tile_diagram} \vspace{-1.2em}
\end{figure}
The FPGA architecture designed provides a highly configurable SIMD engine.  SIMD structures exist in the DNN, and can be mapped to the Processing Elements (PE) present in our architecture.  The accelerator is a scalable SIMD engine, where the data flow is optimized to maximize the number of operations performed for each byte of data fetched.  A layer in ResNet topology can be used many times across the entire topology. Different modules in the hardware must be flexible enough to be re-used across layers, thus helping in keeping the number of hardware resources to a minimum across a large set of resources.

We design a tile based spatial architecture consisting of many tiles and PE's to perform MAC operations, as shown in Figure \ref{fig:System_design}.  Tiles and PE's are easily configurable in our design. The parallelism in convolution is extracted across many output feature maps; within one input feature map, the weights are re-used along all the tiles in PE.  Each PE works on a new set of input pixel points.  As a result, each tile produces different output feature maps. We have fixed the number of tiles to be 64, and the number of PE's per tile to be 4 (based on total compute that can be fit into the chip).

Different hardware components are listed below:

\noindent $\bullet$ Load Store Unit (LSU): Manages data fetch and save from/to system memory through PCIe Avalon interface.  LSU handles multiple outstanding requests.

\noindent $\bullet$ Tile:  Each tile contains an 1-D array of PE's.  This can be parameterizable.

\noindent $\bullet$ PE:   Mapped to ALM's in FPGA, it performs the core compute.

\noindent $\bullet$ IRAM Buffer:     Mapped to block RAMS in FPGA, they are used to store IFMs/inputs.

\noindent $\bullet$ BSRAM Buffer:   Mapped to block RAMS in FPGA, they are used to store Kernels/Weights.

\noindent $\bullet$ SSRAM Buffer:   Mapped to block RAMS in FPGA, they are used to store scaling values.

\noindent $\bullet$ BBSRAM Buffer: Mapped to block RAMS in FPGA, they are used to store Bias values.

\noindent $\bullet$ ORAM Buffer:    Mapped to block RAMS in FPGA, they are used to store OFMs/outputs.

\noindent $\bullet$ Element\_Wise Buffer: Mapped to block RAMS in FPGA, they are used to bring in previously computed OFM data from memory for element-wise addition. 

\begin{figure}[h] \begin{center}
 \includegraphics[scale=0.45]{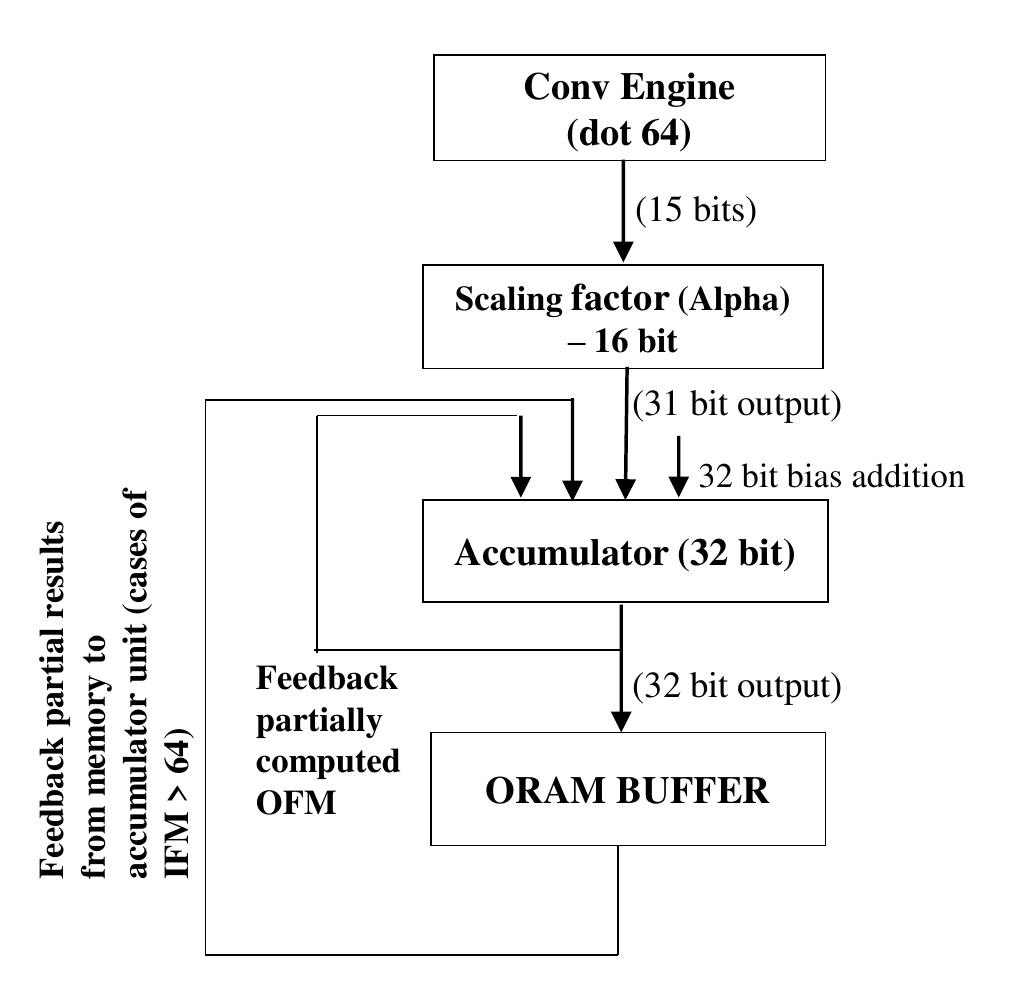}
 \end{center} \vspace{-3em}
 \caption{Core compute engine pipeline}
  \label{fig:core_engine} \vspace{-.90em}
\end{figure}

 \vspace{-1.0em}
\subsection{Core Execution Pipeline}
Our core compute engine pipeline, shown in Figure \ref{fig:core_engine}, is designed with deep pipeline stages (20).  Different computing models present inside the PE are shown Figure \ref{fig:core_engine}.  All the computing elements are mapped into ALM's.

\noindent $\bullet$  Dot64 Engine - Tile layout of our architecture is shown in Figure \ref{fig:Tile_diagram}.  Each PE performs a dot64 compute.  A single dot64 engine block performs MAC computations on 64 pixels of IFM, and 64 pixels of weights that result in a 15-bit output.  Ternary multiplication operation is executed with LUT based multiplier without the need for actual multiplier circuitry.  If the weights are -1, the computation can be simplified by negating the input value.  Each PE in the tile produces one pixel of the output feature maps (OFMs).  A single dot64 operation is optimized and mapped into ALM's.  A single dot64 logic uses 660 ALM's and operates at 660MHz with high packing efficiency.  Optimizing the PE logic design is critical since peak performance is directly related to the number of PE logic blocks that we can fit in our tile based logic.

\noindent $\bullet$  Scaling Engine - Output of dot64 is multiplied by a 16-bit scaling weight, resulting in a 31-bit output.

\noindent $\bullet$  Accumulator/Bias unit - Each partially computed output pixel is accumulated (32-bit accumulator), resulting in a 32-bit pixel output.  The bias value is also added to each of the computed output pixel values.  The partially computed pixel values are then fed back from the accumulator into the accumulator engine in the next cycle until a full OFM is computed.  After the full OFM is computed, the value is written into the OSRAM Buffer. 

Our architecture is fully pipelined so that we feed new data into different compute elements of the PE every cycle.  There are 4 read channels, which read 64 bytes of data for input, weights, scaling values and bias.  They store the read values into the internal memory.  The data is read from the internal memory, and fed into the compute engine.  Memory controllers are responsible for requesting new input data, filling the memory buffers, draining the sections of output completed and generating request to system memory to write the data.

 We have one output write channel to write the computed OFM pixel values into the memory.  
%In ResNet topology, IFM and OFM sizes scale in multiples of 64. 
If the layer consists of 128 IFM's and 64 OFM's, the first set of 64 IFM's and 64 OFM's are computed and stored in the OSRAM Buffer.  The next set of 64 IFM's are read from memory, passed into the compute engine.  During the accumulator stage, we read the previously computed partial outputs (from the previous 64 IFM's) from memory, accumulate with the currently computed value, and store the computed value back into memory.

\begin{figure}[h] \begin{center}
 \includegraphics[scale=0.45]{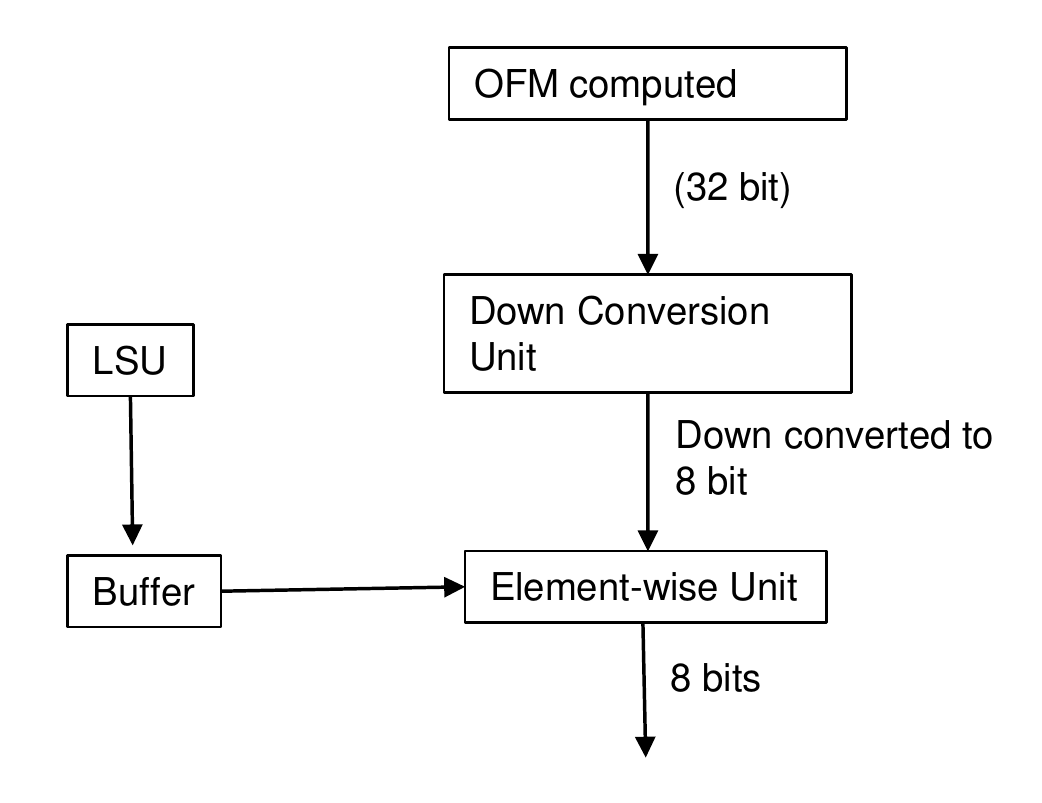}
 \end{center} \vspace{-3em}
 \caption{Down conversion flow with Element-Wise unit}
  \label{fig:DC_flow} \vspace{-1.0em}
\end{figure}

DFP data format used in this work consists of a combination of integer, $I$, and shared exponent, $E_s$.  In this work, we use a single shared exponent (8-bits) per layer for weights and inputs.  Once all the OFM values are computed, we down-convert the 32-bit output pixel values to 8-bits, as outlined below.

\noindent $\bullet$ We find the absolute (abs) max from all the OFM values. 

\noindent $\bullet$ From the max value, we determine the shift value for down conversion. The shift value is determined by using a Leading Zero Count (LZC) detector, as shown in the expression given below. P represents the number of bits used by integer elements in the OFM. 
\begin{align}\label{eq:dfp_downconvert}
\begin{split}
	 % R_s = P - LZC(\max\limits_{\forall i_{ab} \in I^{32}} |i_{ab}|) \\
           R_s = P - LZC(\max\limits_{\forall ofm \in I^{32}} |ofm|) \\
          ofm^{d} = ofm \textit{>}\textit{>} R_s \textit{ and exponent, } 
          E_s^{ab} += R_s
\end{split}
\end{align}

\noindent $\bullet$ The same shift value will be used across all the OFM pixel points.

\begin{figure}\begin{center}
 \includegraphics[scale=0.45]{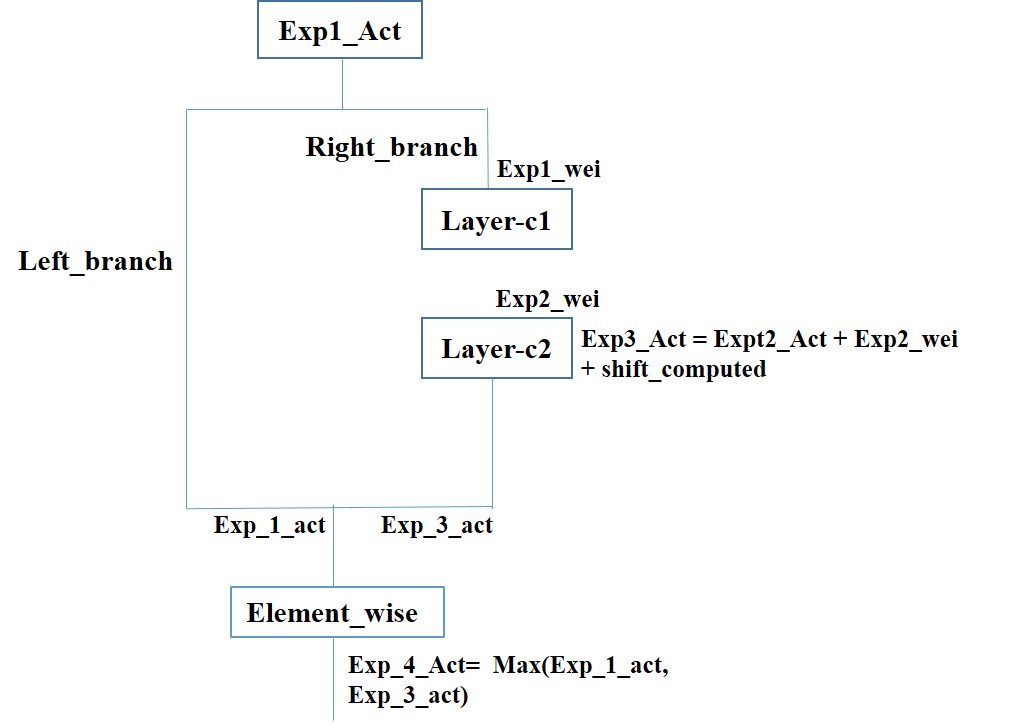}
 \end{center} \vspace{-2em}
 \caption{Exponent computation flow}
  \label{fig:Exp_flow} \vspace{-2.0em}
\end{figure}

\noindent $\bullet$ With the shift value determined ($R_s$), we right shift, and down convert to 8 bits ($ofm^d$).  Then, we round depending on the value of the round and bias bits.  The first two bits after the right shift are the round and bias bits.  the rest of the seven bits along with the sign bit are our down-converted bits.  If both the bias and round bits are not set to 0, we add 1 to our down-converted output.  Exponent activation for the layer is computed by summing the current exponent for activation, the weight exponent, and the value of the down conversion shift computed previously, as shown in Equation \ref{eq:dfp_downconvert} and Figure \ref{fig:Exp_flow}.  Total exponent value computed is passed to the next layer, which becomes the activation exponent for the next layer.

The down-converted output is written back to memory thru the LSU.  We combine the first 8-bit pixel from each of the 64 OFM's (from 64 tiles), which results in one cache line, and write that block of data to the memory.  Correspondingly, this is done for the other OFM pixels.

%As our architecture works on DPF (dynamic fixed point), activation and weights also have exponent values associated with them. This architecture contains one exponent for the weights per layer and exponent for the activations are computed dynamically as shown in Figure 5. Exponent activation for the layer is computed by summing up the current exponent for activation with the weight exponent and value of down conversion shift value computed previously. Total activation exponent computed is passed as input to next layer (activation exponent).
%The exponent values are read (exponent –weights) from the registers to compute the total exponent for the particular layer. Once the exponent values are computed, they are written back into register. The exponent values computed are also passed into the element-wise block.

ResNet also has element-wise layers, where the left and right convolution branches are merged, and added together to get an 8-bit value.  At any moment of time, our accelerator can process a left or right branch of ResNet topology.  For example, left branch convolutions are performed, and the outputs are stored in the memory.  Next the right branch of convolutions are executed in the pipeline.  The OFM computed is stored in the OSRAM Buffer (32 bits each).  We then read the first 32-bit pixel from each of the 64 tiles (32x64), and feed it into the down-conversion logic block.  We obtain the 8-bit down-converted output as explained above (8x64), resulting in a total of 512 bits.  When the element-wise layer is initiated, we read the previously computed values from the LSU into the buffer.  At every cycle, we feed 512 bits (8x64) of data from the buffer into the element-wise logic unit.  The data is read from the buffer. 
 
%As our architecture works on DFP (dynamic fixed point), activation and weights also have exponent values associated with them. There is one exponent for the weights per layer and exponent for the activations are computed dynamically as shown in Figure \ref{fig:Exp_flow}. The activation exponent for the layer is computed by summing up the current exponent for activation with the weight exponent and value of shift value computed previously (shift determined for down conversion). Total activation exponent computed is passed as input to next layer (activation exponent).

Adding two 8-bit DFP's produces an 8-bit output, and a new shared exponent.  In case of the element-wise layer, we add the output of two convolution layers ($ofm_a$/$ofm_b$) together to produce an 8-bit output.  Since the left and right branches have different exponent values computed, we perform a shift (finding the greater of the two exponents), and then add them in the element-wise unit block.  This results in 8-bit values, as shown in Equation \ref{eq:dfp_add}.  $E_s^a$ and $E_s^b$ are the shared exponents of the left and right branch layers.

\begin{equation}\label{eq:dfp_add}
	ofm_{a+b} = \begin{cases}
    		  	ofm_a + ofm_b \textit{>}\textit{>} (E_s^a-E_s^b) \textit{, when }E_s^a \textit{>} E_s^b \\
    		  	ofm_b + ofm_a \textit{>}\textit{>} (E_s^b-E_s^a) \textit{, when }E_s^b \textit{>} E_s^a 
			\end{cases} 
 \end{equation}

\iffalse
 
\begin{itemize}
\item Multiplying two DFP-16 tensors produces 32-bit $I$ tensor with a new shared exponent expressed as follows.  
\begin{align}\label{eq:dfp_multiply}
\begin{split}
      i_{ab} = i_a \times i_b  \textit{ and exponent, } 
      E_s^{ab} = E_s^a + E_s^b
\end{split}
\end{align}
\item Adding two DFP-16 tensors may result in a 32-bit $I$ tensor and a new shared exponent.
\begin{equation}\label{eq:dfp_add}
	i_{a+b} = \begin{cases}
    		  	i_a + i_b \textit{>}\textit{>} (E_s^a-E_s^b) \textit{, when }E_s^a \textit{>} E_s^b \\
    		  	i_b + i_a \textit{>}\textit{>} (E_s^b-E_s^a) \textit{, when }E_s^b \textit{>} E_s^a 
			\end{cases}
            
\end{equation}
Note that when a Fused Multiply and Add operation is performed, all products have the same shared exponent: $E_s^{ab} = E_s^a + E_s^b$, and hence the sum of such products also has the same shared exponent. 
\item Down-Conversion scales DFP-32 output of a layer to DFP-16 to be passed as input to the next layer. The 32-bit $I$ tensor right-shifted $R_s$ bits to fit into 16-bit tensor. The $R_s$ value and the new shared exponent are expressed as follows.  
\begin{align}\label{eq:dfp_downconvert}
\begin{split}
	  R_s = P - LZC(\max\limits_{\forall i_{ab} \in I^{32}} |i_{ab}|) \\
      i_{ab}^d = i_{ab} \textit{>}\textit{>} R_s \textit{ and exponent, } 
      E_s^{ab} += R_s
\end{split}
\end{align}
\end{itemize}
\fi

 \vspace{-1.3em} 
\section{Memory Organization/Layer Programming} \label{sec:Memory Organization}
Based on the requirements of each layer, we construct a memory layout for each of the inputs, weights, and scaling values for efficient data distribution into different memory buffers.  The LSU controller helps in movement of data between the memory and on-chip buffers.

\noindent $\bullet$ BSRAM : Kernal memory layout is arranged in system memory by combining each of the 2-bit pixels from 64 weights.  Each pixel in the kernal (3x3) will be 128 bits.  Kernal data will be then distributed to 64 tiles. BSRAM buffer holds the weight values.
In our design with 64 tiles, there is a separate BSRAM buffer per tile to feed the data into the PE's present in all the tiles.  Distribute logic in the BSRAM distributes the data that is read from the memory into individual BSRAM buffers present in each of the tiles.  To reduce the load on the distribute logic, we divide it into set of four distribute logic blocks, which feed a set of 16 tiles.  Each BSRAM buffer has a controller, which initiates the reading and writing into the buffer.
 
\noindent $\bullet$ SSRAM: Scaling memory layout is arranged in a similar manner to the BSRAM, where each scaling value is 16 bits. SSRAM buffer holds the scaling value.  There is a separate SSRAM buffer per tile to feed the scaling values into each of the PE's present on the tiles.  The SSRAM distribute logic distributes the data read from memory to the individual SSRAM buffer, where we divide it into a set of two distribute logic blocks that feed a set of 32 tiles.

\noindent $\bullet$ BBSRAM: Bias memory layout contains one 32-bit value depending on number of the OFMs.  The BBSRAM buffer holds the bias value.  We add them to each of the computed OFM pixel points.

\noindent $\bullet$ ISRAM: ISRAM memory layout is organized in system memory by combining each of the 8-bit pixels from the 64 IFM's (combine along z-depth).  Every pixel in our modified input image would be 512 bits.  Internally, the 512 bits would be read, and distributed to each of the PE's on the tiles.  ISRAM logic consists of 1 bank per PE (4 PE's per tile); the ISRAM controller the feeds data needed for computation into the corresponding PE based on the stride information.
%\section{Layer Programming/Sequential Opertion} \label{sec:Layer Programming/Sequential operation}

ResNet contains a sequence of layers that need to be executed sequentially in our hardware.  Based on the topology of the particular layer, our accelerator contains a set of registers that need to be programmed for DNN execution.  Our register programming is divided into 2 parts:\\
$\noindent$1. Program the core registers: Provide the layer topology information through the registers, such as input, output feature map dimensions, kernal size, stride, number of channels, and number of tiles that are stored in the configuration register. \\
$\noindent$2. Program the LSU registers: Provide data fetch information (size of the data that needs to be fetched, and the base address) through these registers.  These set of registers will provide information on the quantity of IFM and kernel data that need to be fetched from memory, and OFM data to be written to memory.

The control logic keeps track of the current number of executed layers, and loads the corresponding values into the registers for each layer upon  completion of the current layer (transfer of data from layer 1 to layer 2).

\vspace{-1.0em}
\section{Experimental Results}\label{sec:Experimental Results}
Our proposed DNN hardware accelerator for ResNet50 topology was implemented on two of Intel's state-of-the-art FPGA's, the Arria10 GX1150 and the Stratix10.  The complete DNN accelerator was written in RTL.   We used Intel Quartus Prime software, and EPE tool to estimate performance and power.  For the Stratix10, we used Quartus Early Beta Release to evaluate our network. 

We retrained the ResNet50 ternary model, using the fine tuning method described in \cite{TernaryFGQ}. We optimized this model to suit the design parameters of the hardware and fused the batchnorm paramters into convolution layers. We validated the accuracy of the fused model using a modified version of Caffe with support for emulated ternary operations and acheived top-1 accuracy of 71.1\% for ResNet50.  

We began our experiment by running convolutions with 8-bit activations and ternary weights on the Arria10 FPGA. The first and last layers of ResNet50 are executed on the CPU(8-8)  which account for only 4.3\% of the total flops (3.8GMACs).
The Altera Aria10 GX1150 contains 427K ALM's with 2713 M20K RAM blocks.  The different RAM blocks in our design shown in Table \ref{table:Ram} were mapped to M20 blocks in FPGA.
Table \ref{table:Arria10} shows the FPGA resource utilization on implementing a ResNet network on the Arria10 FPGA. 
Total ALM usage was 83\% for mapping the entire ResNet with DFP support onto the Arria10. 
In Table \ref{table:ALM}, we provide the breakdown of ALM usage for one instance of the different logic/memory modules present in the design.  Some of the modules were instantiated many times in the design
\vspace*{-\baselineskip}
\begin{table}[h]
\centering \vspace{-1em}
\caption{Arria10 Resource Utilization}
\vspace{0.5em}
\begin{tabular}{|c|c|} \hline
Precision      & 8-2     \\ \hline
ALM usage      & 83\%    \\ \hline
DSP usage      & 0       \\ \hline
Freq(MHz)      & 200     \\ \hline
Accuracy(top1) & 71.10\% \\ \hline
RAM(M20K)      & 19\%    \\ \hline
Dot operation  & 64  \\ \hline    
\end{tabular}
\label{table:Arria10}
\vspace{-.25em}
\end{table}

\vspace*{-\baselineskip}
\begin{table}[h]
\centering
\caption{Resource usage of different buffers}
\begin{tabular}{|c|c|c|c|}
\hline
Buffer & Dimension & Instances & \begin{tabular}[c]{@{}c@{}}Memory\\ (MB)\end{tabular} \\ \hline
IRAM & 512 X 128 & 8 & 0.524 \\ \hline
BSRAM & 128 X 128 & 64 & 0.13 \\ \hline
ORAM & 128 X 1028 & 64 & 1.29 \\ \hline
BBSRAM & 32x128 & 64 & 0.2445 \\ \hline
ElementWise & 512*128 & 1 & 0.065 \\ \hline
SSRAM & 16 X128 & 64 & 0.0163 \\ \hline
\end{tabular}
\label{table:Ram}
\vspace{-1em}
\end{table}

\begin{table}[h]
\centering \vspace{-1em}
\caption{ALM usage for diifferent Modules}
\begin{tabular}{|c|c|}
\hline
Modules & ALM Usage \\ \hline
Dot64 & 660 \\ \hline
Load Unit & 2143 \\ \hline
Store Unit & 342 \\ \hline
DownConversion & 108 \\ \hline
ElementWise & 36 \\ \hline
Memory\_ctrl & 4476 \\ \hline
Memory\_dist.logic & 7510 \\ \hline
Max\_logic & 242 \\ \hline
Avalon\_interface & 100 \\ \hline
\end{tabular}
\label{table:ALM}
\vspace{-1.5em}
\end{table}

\begin{figure}[h] 
\begin{center} \vspace{-0.75em}
 \includegraphics[scale=0.45]{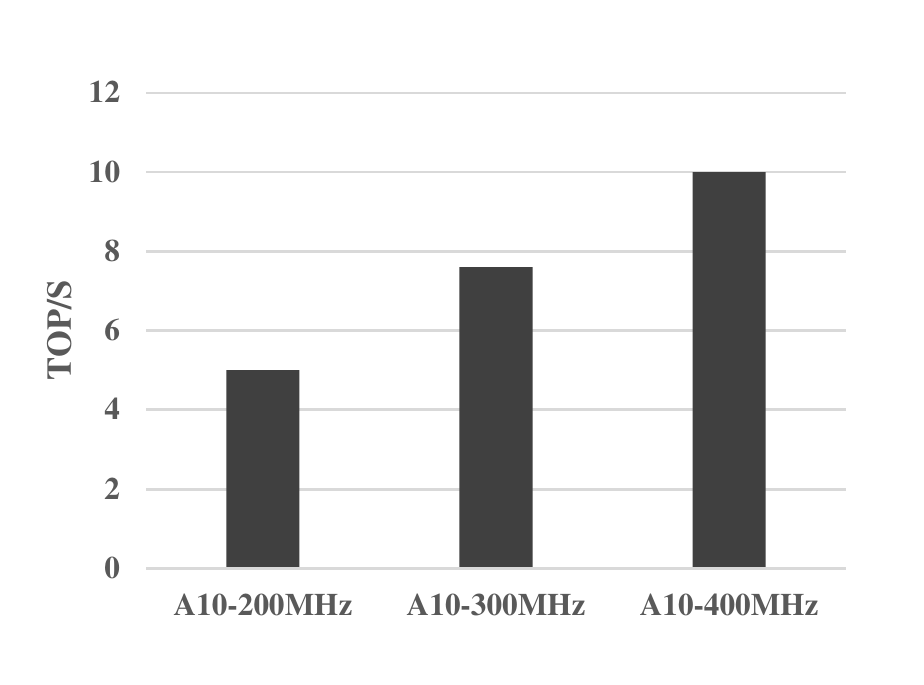}
 \end{center} \vspace{-2.5em}
 \caption{Tops obtained on A10 for ResNet50 network at varying aggressive frequency targets.}
  \label{fig:A10_Tops} \vspace{-1em}
\end{figure}

\begin{figure}[h]
\begin{center} \vspace{-0.5em}
 \includegraphics[scale=0.45]{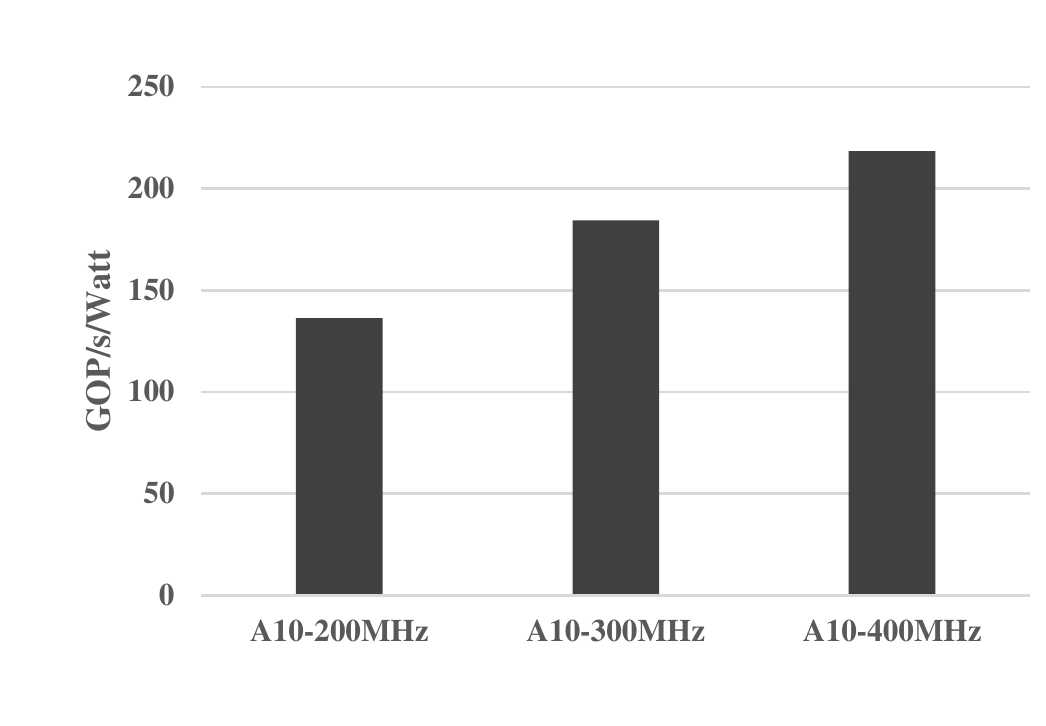}
 \end{center} \vspace{-3em}
 \caption{GOP/s/Watt obtained on A10 for ResNet50 network at varying aggressive frequency targets.}
  \label{fig:power_A10_Tops} \vspace{-1em}
\end{figure}

We compared the performance obtained for ResNet50 on the Arria10 and the Stratix10. We believe our design can be further optimized and thus we make performance projections with more aggressive performance targets (300MHz and 400MHz). Figures \ref{fig:A10_Tops} and \ref{fig:power_A10_Tops} show TOP/s and GOP/s/Watt obtained for ResNet50 on the Arria10. 
Our design obtained 5 TOP/s on ResNet50. This is the highest performance number reported on the Arria10 for ResNet50.
With an aggressive performance target of 400MHZ, which we believe is achievable, we can obtain performance of 10TOP/s. 

\begin{figure}[h] 
\begin{center} \vspace{-.25em}
 \includegraphics[scale=0.45]{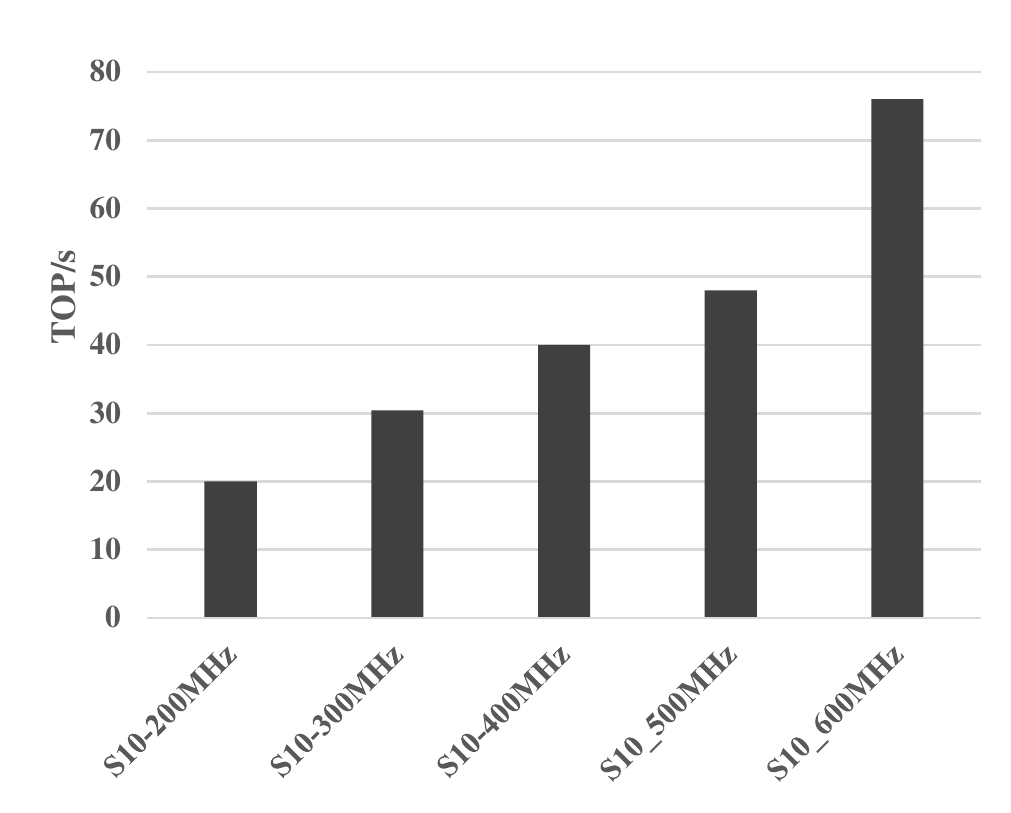}
 \end{center} \vspace{-2em}
 \caption{Peak Tops obtained on S10 for ResNet50 network at varying aggressive frequency targets.}
  \label{fig:S10_Tops} \vspace{-.95em}
\end{figure}

\begin{figure}[h]
\begin{center} 
 \includegraphics[scale=0.45]{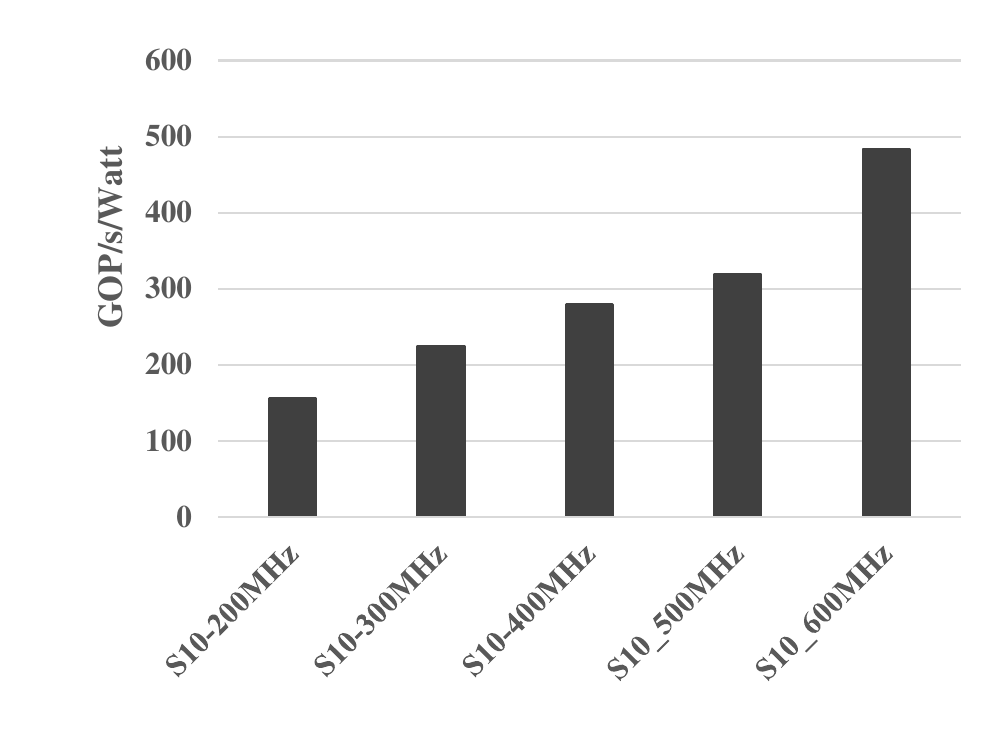}
 \end{center} \vspace{-3em}
 \caption{GOP/s/Watt obtained on S10 for ResNet50 network at varying aggressive frequency targets.}
  \label{fig:power_S10_Tops} \vspace{-.95em}
\end{figure}

Stratix10 results are obtained for the S10\_SG280 part (933K ALM's) using Quartus tools. By studying performance on the lower-end S10 part, we project performance on the Stratix10's highest-end part (S10\_SG550), which has 1.8M ALM's .  The Stratix10 is the advanced FPGA device with HyperFlex support (Hyperflex can enable frequency > 600MHz). On the S10 (SG550), we project a performance of 76 TOP/s and 0.78 TOP/s/Watt for ResNet50, as shown in Figures \ref{fig:S10_Tops}and \ref{fig:power_S10_Tops}.

%\vspace*{-\baselineskip}

\begin{figure}[h]
\begin{center}
 \includegraphics[scale=0.45]{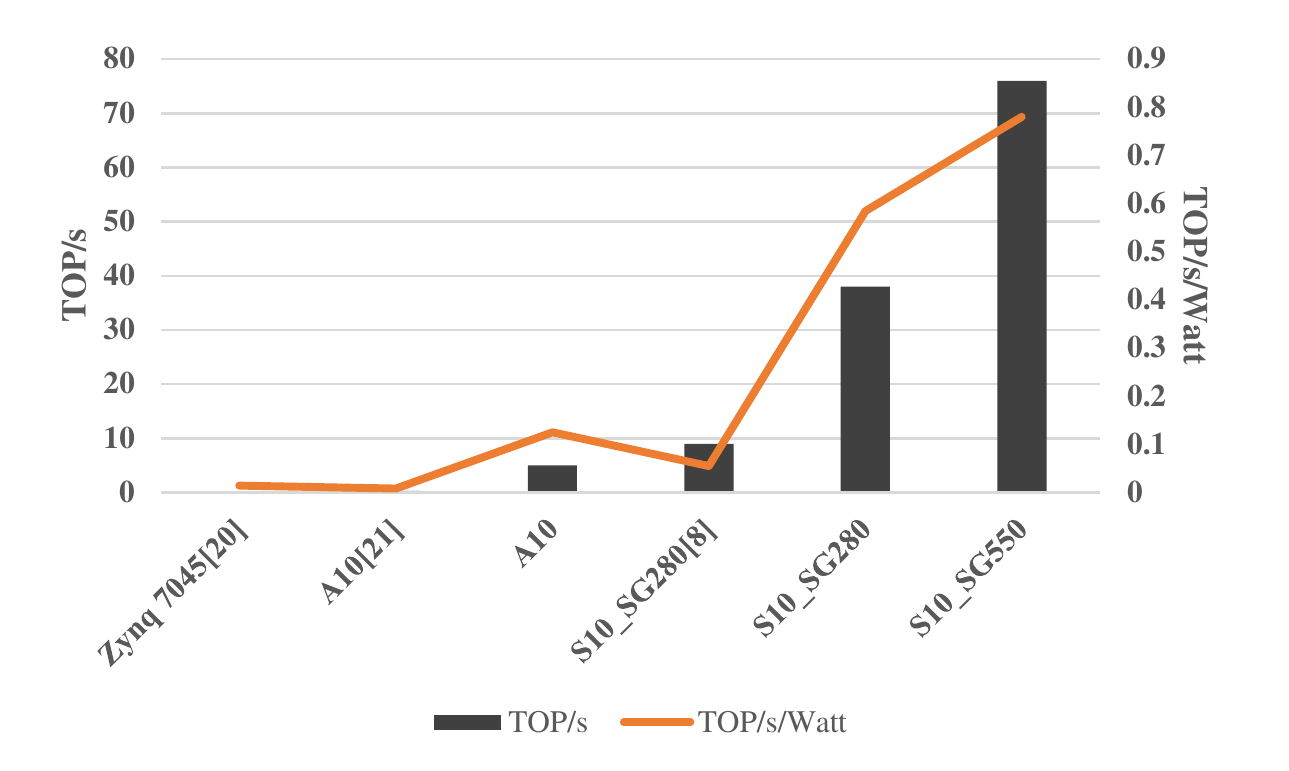}
 \end{center} \vspace{-3em}
 \caption{Performance/Watt comparison of our ResNet 50 results on A10/S10 with other prior work.}
  \label{fig:FPGA_compare} \vspace{-.95em}
\end{figure}

There has been numerous research on DNN implementation on FPGA. However, very few works have shown FPGA performance on state-of-the-art topologies, like ResNet50, with ternary weights and on latest Intel FPGA. We compare our S10\_SG280 performance estimates for ResNet50 with results reported in \cite{Nurvitadhi:2017} for the same S10 part. The ternary ResNet reported in  \cite{Nurvitadhi:2017} uses 2-bit weights and 32-bit activations, which is different from 8-2 presented in this work. Our S10 performance numbers show >4x improvement in TOP/s compared to \cite{Nurvitadhi:2017}, as shown in Figure  \ref{fig:FPGA_compare}. High performance is achievable as lower precision width uses less logic. Thus, this frees up space to pack more compute into our design. We also extract more performance from our design by fine tuning the dot module without using any DSP's. This results in lower ALMs/op and achieves high packing efficiency.
 
We also compare the TOP/s obtained for ResNet50 on S10/A10 with other architectures such as the Arria10 and Xilink Zynq Z-7045 SOC in Figure \ref{fig:FPGA_compare}. ResNet50 implementation on the Arria10 with 16-bit precision (A10 [20]) was shown to achieve 0.315 TOP/s \cite{ResNet}. Their entire design used 69\% DSP's and 30\% ALM's.  By lowering the precision, we are able to use ALM's in our design effectively and 
achieve more than 15x performance improvement on the A10 compared to \cite{ResNet}.
Recent ResNet50 implementation on Xilink Zynq Z-7045 was shown to achieve 0.128TOP/s \cite{Snowflake}. This design used 16-bit fixed precision and can execute 256 MAC/cycle. Our design can perform 16K MAC/cycle and our S10/A10 performance outdoes the Xilink comprehensively, as shown in Figure \ref{fig:FPGA_compare}.

We also compare peak TOP/s obtained with the Titan X GPU for Ternary ResNet50 (32(a)-2(w)) in \cite{Nurvitadhi:2017}. The S10 (SG280) outperforms the Titan X GPU by more than 5x in performance. GPU's are not yet designed to effectively implement low precision operations and thus FPGA's outperform the GPU in performance achieved. 

\begin{table}[h]
\centering \vspace{-1em}
\caption{Theoretical Peak TOP/s and TOP/s/Watt of recent NVIDIA GPU for Inferencing \cite{Nvidia}}
\label{table:GPU}
\begin{tabular}{|c|c|c|c|c|}
\hline
GPU  & TOP/s & \begin{tabular}[c]{@{}c@{}}Power\\ (W)\end{tabular} & Precision & TOP/s/Watt \\ \hline
V100 & 125   & 300                                                 & FP16      & 0.416667   \\ \hline
P4   & 22    & 75                                                  & INT8      & 0.293333   \\ \hline
P40  & 47    & 250                                                 & INT8      & 0.188      \\ \hline
P100 & 18.7  & 250                                                 & FP16      & 0.0748     \\ \hline
\end{tabular} \vspace{-1em}
\end{table}

Various recent NVIDIA GPU offerings for Inferencing are shown in Table \ref{table:GPU}. Our A10 and S10 (SG\_280/SG\_550) platforms can achieve better performance/watt for ResNet50 compared to theoretical TOP/s/Watt offered by these GPU devices.

\vspace{-1em}
\section{Conclusions} \label{sec:Conclusions}
In this paper, we have designed a high performance DNN accelerator that can accelerate low precision inference for deep learning.  There is trade-off between being able to accomodate more compute into hardware using low precision data types and retaining accuracy compared to full precision.  In this work, with fused a ResNet50 network and applied FGQ  technique to ternarize the weight.  We achieved a top-1 accuracy of 71.1\%.  We have efficiently implemented a ResNet50 network with 8-bit activations and ternary weights using only ALM's on FPGA. We have optimized our core compute engine such that we can efficiently pack more compute into a FPGA. Our design is capable of performing 16K MAC operations per cycle. Our experimental results indicated that performance achieved by our design on the Arria10 and Stratix10 for ResNet50 can out-perform other hardware implementations of the ResNet50 on FPGA's such as the Arria10, Xilink, and state-of-the-art GPU.

\vspace{-1em}

%\end{document}  % This is where a 'short' article might terminate

%ACKNOWLEDGMENTS are optional
%\section{Acknowledgments}

%
% The following two commands are all you need in the
% initial runs of your .tex file to
% produce the bibliography for the citations in your paper.
%\bibliographystyle{abbrv}

\renewcommand*{\bibfont}{\small}

\bibliographystyle{plainnatedited}
 \setlength{\bibsep}{0pt plus 0.3ex} 
%\bibliography{references} 

% sigproc.bib is the name of the Bibliography in this case
%\bibliography{references}
% You must have a proper ".bib" file
%  and remember to run:
% latex bibtex latex latex
% to resolve all references
%
% ACM needs 'a single self-contained file'!
%
%APPENDICES are optional
%\balancecolumns
%\appendix
%Appendix A
\end{document}